
%
%
\input harvmac %
%
%
%
%
%
%
%
%
%
\newif\ifdraft

\noblackbox
\catcode`\@=11
\newif\iffrontpage
%
\ifx\answ\bigans
\def\titleft{\titsm}
\magnification=1200\baselineskip=15pt plus 2pt minus 1pt
%
\advance\hoffset by-0.075truein
\hsize=6.15truein\vsize=600.truept\hsbody=\hsize\hstitle=\hsize
\else\let\lr=L
\def\titleft{\titla}
\magnification=1000\baselineskip=14pt plus 2pt minus 1pt
%
\vsize=6.5truein
\hstitle=8truein\hsbody=4.75truein
\fullhsize=10truein\hsize=\hsbody
\fi
\parskip=4pt plus 10pt minus 4pt

\font\titla=cmr10 scaled\magstep3
\font\tenmss=cmss10
\font\absmss=cmss10 scaled\magstep1
\newfam\mssfam
\font\footrm=cmr8  \font\footrms=cmr5
\font\footrmss=cmr5   \font\footi=cmmi8
\font\footis=cmmi5   \font\footiss=cmmi5
\font\footsy=cmsy8   \font\footsys=cmsy5
\font\footsyss=cmsy5   \font\footbf=cmbx8
\font\footmss=cmss8
\def\footfont{\def\rm{\fam0\footrm}
\textfont0=\footrm \scriptfont0=\footrms
\scriptscriptfont0=\footrmss
\textfont1=\footi \scriptfont1=\footis
\scriptscriptfont1=\footiss
\textfont2=\footsy \scriptfont2=\footsys
\scriptscriptfont2=\footsyss
\textfont\itfam=\footi \def\it{\fam\itfam\footi}
\textfont\mssfam=\footmss \def\mss{\fam\mssfam\footmss}
\textfont\bffam=\footbf \def\bf{\fam\bffam\footbf} \rm}
\def\tenpoint{\def\rm{\fam0\tenrm}
\textfont0=\tenrm \scriptfont0=\sevenrm
\scriptscriptfont0=\fiverm
\textfont1=\teni  \scriptfont1=\seveni
\scriptscriptfont1=\fivei
\textfont2=\tensy \scriptfont2=\sevensy
\scriptscriptfont2=\fivesy
\textfont\itfam=\tenit \def\it{\fam\itfam\tenit}
\textfont\mssfam=\tenmss \def\mss{\fam\mssfam\tenmss}
\textfont\bffam=\tenbf \def\bf{\fam\bffam\tenbf} \rm}
\ifx\answ\bigans\def\abstractfont{\tenpoint}\else
\def\abstractfont{\def\rm{\fam0\absrm}
\textfont0=\absrm \scriptfont0=\absrms
\scriptscriptfont0=\absrmss
\textfont1=\absi \scriptfont1=\absis
\scriptscriptfont1=\absiss
\textfont2=\abssy \scriptfont2=\abssys
\scriptscriptfont2=\abssyss
\textfont\itfam=\bigit \def\it{\fam\itfam\bigit}
\textfont\mssfam=\absmss \def\mss{\fam\mssfam\absmss}
\textfont\bffam=\absbf \def\bf{\fam\bffam\absbf}\rm}\fi
%
\def\f@@t{\baselineskip10pt\lineskip0pt\lineskiplimit0pt
\bgroup\aftergroup\@foot\let\next}
\setbox\strutbox=\hbox{\vrule height 8.pt depth 3.5pt width\z@}
\def\vfootnote#1{\insert\footins\bgroup
\baselineskip10pt\footfont
\interlinepenalty=\interfootnotelinepenalty
\floatingpenalty=20000
\splittopskip=\ht\strutbox \boxmaxdepth=\dp\strutbox
\leftskip=24pt \rightskip=\z@skip
\parindent=12pt \parfillskip=0pt plus 1fil
\spaceskip=\z@skip \xspaceskip=\z@skip
\Textindent{$#1$}\footstrut\futurelet\next\fo@t}
\def\Textindent#1{\noindent\llap{#1\enspace}\ignorespaces}
\def\footnote#1{\attach{#1}\vfootnote{#1}}%

\def\foot{\attach\footsymbolgen\vfootnote{\footsymbol}}
\let\footsymbol=\star
\newcount\lastf@@t           \lastf@@t=-1
\newcount\footsymbolcount    \footsymbolcount=0
\def\footsymbolgen{\relax\footsym
\global\lastf@@t=\pageno\footsymbol}
\def\footsym{\ifnum\footsymbolcount<0
\global\footsymbolcount=0\fi
{\iffrontpage \else \advance\lastf@@t by 1 \fi
\ifnum\lastf@@t<\pageno \global\footsymbolcount=0
\else \global\advance\footsymbolcount by 1 \fi }
\ifcase\footsymbolcount \fd@f\star\or
\fd@f\dagger\or \fd@f\ast\or
\fd@f\ddagger\or \fd@f\natural\or
\fd@f\diamond\or \fd@f\bullet\or
\fd@f\nabla\else \fd@f\dagger
\global\footsymbolcount=0 \fi }
\def\fd@f#1{\xdef\footsymbol{#1}}
\def\space@ver#1{\let\@sf=\empty \ifmmode #1\else \ifhmode
\edef\@sf{\spacefactor=\the\spacefactor}
\unskip${}#1$\relax\fi\fi}
\def\attach#1{\space@ver{\strut^{\mkern 2mu #1}}\@sf}
%
\newif\ifnref
\def\rrr#1#2{\relax\ifnref\nref#1{#2}\else\ref#1{#2}\fi}
\def\ldf#1#2{\begingroup\obeylines
\gdef#1{\rrr{#1}{#2}}\endgroup\unskip}
\def\nrf#1{\nreftrue{#1}\nreffalse}
\def\doubref#1#2{\refs{{#1},{#2}}}

\nreffalse
\def\refout{\listrefs}
%
\def\eqn#1{\xdef #1{(\secsym\the\meqno)}
\writedef{#1\leftbracket#1}%
\global\advance\meqno by1\eqno#1\eqlabeL#1}
\def\eqnalign#1{\xdef #1{(\secsym\the\meqno)}
\writedef{#1\leftbracket#1}%
\global\advance\meqno by1#1\eqlabeL{#1}}
%
\def\chap#1{\newsec{#1}}
\def\chapter#1{\chap{#1}}
\def\sect#1{\subsec{{ #1}}}
\def\section#1{\sect{#1}}
\def\\{\ifnum\lastpenalty=-10000\relax
\else\hfil\penalty-10000\fi\ignorespaces}
\def\note#1{\leavevmode%
\edef\@@marginsf{\spacefactor=\the\spacefactor\relax}%
\ifdraft\strut\vadjust{%
\hbox to0pt{\hskip\hsize%
\ifx\answ\bigans\hskip.1in\else\hskip-.1in\fi%
\vbox to0pt{\vskip-\dp
\strutbox\sevenbf\baselineskip=8pt plus 1pt minus 1pt%
\ifx\answ\bigans\hsize=.7in\else\hsize=.35in\fi%
\tolerance=5000 \hbadness=5000%
\leftskip=0pt \rightskip=0pt \everypar={}%
\raggedright\parskip=0pt \parindent=0pt%
\vskip-\ht\strutbox\noindent\strut#1\par%
\vss}\hss}}\fi\@@marginsf\kern-.01cm}
\def\titlepage{%
\frontpagetrue\nopagenumbers\abstractfont%
\hsize=\hstitle\rightline{\vbox{\baselineskip=10pt%
{\abstractfont\pubnum}}}\pageno=0}
\frontpagefalse
\def\pubnum{}
\def\pdate{\number\month/\number\yearltd}
\def\makefootline{\iffrontpage\vskip .27truein
\line{\the\footline}
\vskip -.1truein\leftline{\vbox{\baselineskip=10pt%
{\abstractfont\pdate}}}
\else\vskip.5cm\line{\hss \tenrm $-$ \folio\ $-$ \hss}\fi}
\def\title#1{\vskip .7truecm\titlestyle{\titleft #1}}
\def\titlestyle#1{\par\begingroup \interlinepenalty=9999
\leftskip=0.02\hsize plus 0.23\hsize minus 0.02\hsize
\rightskip=\leftskip \parfillskip=0pt
\hyphenpenalty=9000 \exhyphenpenalty=9000
\tolerance=9999 \pretolerance=9000
\spaceskip=0.333em \xspaceskip=0.5em
\noindent #1\par\endgroup }
\def\autskip{\ifx\answ\bigans\vskip.5truecm\else\vskip.1cm\fi}
\def\author#1{\vskip .7in \centerline{#1}}

\def\address#1{\ifx\answ\bigans\vskip.2truecm
\else\vskip.1cm\fi{\it \centerline{#1}}}
\def\abstract#1{
\vskip .5in\vfil\centerline
{\bf Abstract}\penalty1000
{{\smallskip\ifx\answ\bigans\leftskip 2pc \rightskip 2pc
\else\leftskip 5pc \rightskip 5pc\fi
\noindent\abstractfont \baselineskip=12pt
{#1} \smallskip}}
\penalty-1000}
\def\endpage{\tenpoint\supereject\global\hsize=\hsbody%
\frontpagefalse\footline={\hss\tenrm\folio\hss}}
\def\ack{\goodbreak\vskip2.cm\centerline{{\bf Acknowledgements}}}
\def\CERN{\address{CERN, Geneva, Switzerland}}
\def\bfone{\relax{\rm 1\kern-.35em 1}}
\def\inbar{\vrule height1.5ex width.4pt depth0pt}
\def\IC{\relax\,\hbox{$\inbar\kern-.3em{\mss C}$}}
\def\ID{\relax{\rm I\kern-.18em D}}
\def\IF{\relax{\rm I\kern-.18em F}}
\def\IH{\relax{\rm I\kern-.18em H}}
\def\II{\relax{\rm I\kern-.17em I}}
\def\IN{\relax{\rm I\kern-.18em N}}
\def\IP{\relax{\rm I\kern-.18em P}}
\def\IQ{\relax\,\hbox{$\inbar\kern-.3em{\rm Q}$}}
\def\IR{\relax{\rm I\kern-.18em R}}
\font\cmss=cmss10 \font\cmsss=cmss10 at 7pt
\def\ZZ{\relax\ifmmode\mathchoice
{\hbox{\cmss Z\kern-.4em Z}}{\hbox{\cmss Z\kern-.4em Z}}
{\lower.9pt\hbox{\cmsss Z\kern-.4em Z}}
{\lower1.2pt\hbox{\cmsss Z\kern-.4em Z}}\else{\cmss Z\kern-.4em
Z}\fi}
\def\nup#1({Nucl.\ Phys.\ $\us {B#1}$\ (}
\def\plt#1({Phys.\ Lett.\ $\us  {#1}$\ (}
\def\cmp#1({Comm.\ Math.\ Phys.\ $\us  {#1}$\ (}
\def\prp#1({Phys.\ Rep.\ $\us  {#1}$\ (}
\def\prl#1({Phys.\ Rev.\ Lett.\ $\us  {#1}$\ (}
\def\prv#1({Phys.\ Rev.\ $\us  {#1}$\ (}
\def\mpl#1({Mod.\ Phys.\ \Let.\ $\us  {#1}$\ (}
\def\tit#1|{{\it #1},\ }
%

%

\def\bar{\overline}
\def\us#1{\underline{#1}}

\def\Coe#1.#2.{{#1\over #2}}

\def\coe#1.#2.{\relax{\textstyle {#1 \over #2}}\displaystyle}

\def\notin{\hbox{{$\in$}\kern-.51em\hbox{/}}}

\def\del{\partial}

\catcode`\@=12
%
\def\brk{\hfill\break}
\def\LG{Lan\-dau--Ginz\-burg}
\def\cy{Calabi--Yau}
\def\K{K\"ahler}
\def\pf{Picard--Fuchs}

\def\leel{low-energy effective Lagrangian}
\def\el{effective Lagrangian}
\def\ps{Planck scale}
\def\np{non-perturbative}
\def\ws{weak scale}
\def\re{\mathop{\rm Re}\nolimits}

\def\Da{\Delta_a}

\def\Mpl{M_{\rm Pl}}
\def\Mw{M_{\rm W}}

\def\eff{{\rm eff}}

\def\veff{V_\eff}

\def\del{\partial}

\def\fd{four-dimensional}

%
%
\ldf\ADS{I.~Affleck, M.~Dine and N.~Seiberg, Phys. Rev.
  Lett.  51 (1984) 1026, \nup241 (1984) 493, \nup256 (1985) 557.}
\ldf\ANT{
I.~Antoniadis, K.~Narain and T.~Taylor, \plt B267 (1991) 37.}
\ldf\BFOFW{J.\ Balog, L.\ Feher, L.\ O'Raifeartaigh, P.\ Forga\'cs
and A.\ Wipf, \plt244B (1990) 435; Ann.\ Phys.\ 203 (1990) 194.}
\ldf\BDFS{T.\ Banks,
L.\ Dixon, D.\ Friedan and S.\ Shenker, \nup299 (1988)
 613.}
\ldf\BGa{P. Bin\'etruy and M. K. Gaillard, \nup
B358 (1991) 121.}
\ldf\BGb{P. Bin\'etruy and M. K. Gaillard, \plt   B253
(1990) 119.}
\ldf\BGG{P. Binetruy, G. Girardi and R. Grimm,
LAPP preprint LAPP-TH-337-91.}
\ldf\BV{B. Blok and  A.\ Varchenko, \tit Topological
 conformal field theories and the flat coordinates| preprint
 IASSNS-HEP-91/5.}
\ldf\CF{A.\ Cadavid and S.\ Ferrara, \plt B267 (1991) 193.}
\ldf\Candelas{P.\ Candelas, \nup 298 (1988) 458.}
\ldf\CD{P.\ Candelas and X.C.\ de la Ossa, \nup355 (1991) 455.}
\ldf\CDGP{P.\ Candelas, X.C.\ de la Ossa, P.S.\ Green and
L.\ Parkes, \plt 258B (1991) 118; \nup359 (1991) 21.}
\ldf\CHSW{P.~Candelas, G.~Horowitz,
  A.~Strominger and E.~Witten, \nup258 (1985) 46.}
\ldf\CLO{G. Cardoso Lopes and B. Ovrut, UPR-preprint 0464T }
\ldf\CLMR{
J.~A.~Casas, Z.~Lalak, C.~Mu\~noz and G.G.~Ross,
\nup347 (1990) 243.}
\ldf\Cecotti{S.\ Cecotti,
 \nup 355 (1991) 755, Int.\ J.\ Mod.\ Phys.\ A6 (1991) 1749.}
\ldf\CFG{
S.\ Cecotti, S.\ Ferrara and L.\ Girardello,
Int.\ Mod.\ J.\ Phys.\ A4 (1989) 2475, \plt B213 (1988) 443.}
\ldf\CFV{S. Cecotti, S. Ferrara and M. Villasante, Int. J. Mod.
Phys.  A2 (1987) 1839.}
\ldf\CV{S.\ Cecotti and C.\ Vafa, \nup367 (1991) 359.}
\ldf\DFKZ{J. P. Derendinger, S. Ferrara, C. Kounnas and F. Zwirner,
Cern preprint TH.6004/91}
\ldf\DIN{J.P.~Derendinger, L.E.~Ib\'a\~nez and
  H.P.~Nilles, \plt  155B (1985) 65.}
\ldf\dspert{M.~Dine and
  N.~Seiberg, Phys. Rev. Lett.  57 (1986) 2625.}
\ldf\dines{M.~Dine and N.~Seiberg,
  in {\it Unified String Theories},
  eds. M.~Green and D.~Gross (World Scientific, 1986),
  \plt  162B (1985) 299.}
\ldf\DVV {R.\ Dijkgraaf, E. Verlinde and H. Verlinde,
\nup352(1991) 59.}
\ldf\DIZ  {P.\ Di Francesco, C.\ Itzykson and J.-B.\ Zuber,
\cmp140 (1991) 543.}
\ldf\DRSW{M.~Dine, R.~Rohm, N.~Seiberg and
  E.~Witten, \plt  156B (1985) 55.}
\ldf\DS{M.~Dine and N.~Seiberg, Phys. Rev. Lett.
    55 (1985) 366.}
\ldf\dixontrieste{For a review see
L. Dixon, {\sl in} Proc. of the 1987 ICTP Summer
Workshop in High Energy Physics, Trieste, Italy, ed.~ G.~Furlan,
J.~C.~Pati, D.~W.~Sciama, E.~Sezgin and Q.~Shafi
and references therein.}
\ldf\dixonberkeley{L. Dixon, talk presented at the M.S.R.I. meeting
on
mirror symmetries, Berkeley, May 1991.}
\ldf\dixondpf{L. Dixon, talk presented at the A.P.S. D.P.F.
  Meeting, Houston, 1990, SLAC-PUB 5229.}
\ldf\DFMS{
  S.~Hamidi and C.~Vafa, \nup279 (1987) 465;\brk
  L.~Dixon, D.~Friedan, E.~Martinec and S.~Shenker,
  \nup282 (1987) 13.}
\ldf\DKLa{L.J.\ Dixon, V.S.\ Kaplunovsky and J.\ Louis,
  \nup329 (1990) 27.}
\ldf\DKLb{L.~Dixon, V.~Kaplunovsky and J.~Louis, \nup355 (1991) 649.}
\ldf\DKLP{
L.~Dixon, V.~Kaplunovsky, J.~Louis and M.~Peskin, unpublished.}
\ldf\DS{Drinfel'd and V.~G.~Sokolov,
 Jour.~Sov.~Math. {\bf 30} (1985) 1975.}
\ldf\FFS{S.\ Ferrara,
P.\  Fr\`e and P.\ Soriani, {\it On the moduli space
of the $T^6/Z_3$ orbifold and its modular group}, preprint CERN-TH
6364/92,
SISSA 5/92/EP.}
\ldf\FGN{S. Ferrara, L. Girardello and H. P. Nilles,
\plt  125B (1983) 457}
\ldf\FGPS{S. Ferrara, L.Girardello, O. Piguet and R. Stora,
\plt  B157 (1985) 179.}
\ldf\FKZL{S.~Ferrara, C.~Kounnas, D.~L\"ust and F.~Zwirner,
CERN preprint CERN-TH-6090-91.}
\ldf\FL{S.\ Ferrara and J. Louis, {\it Flat holomorphic
   connections and Picard-Fuchs identities from
   N=2 supergravity}, preprint CERN-TH-6334-91.}
\ldf\FLT{S.\ Ferrara, D.\ L\"ust and S.\ Theisen,  \plt 242B (1990)
39.}
\ldf\FLST{S.~Ferrara, D.~L\"ust, A.~Shapere and
  S.~Theisen, \plt  B225 (1989) 363;
Y.~Park, M.~Srednicki and A.~Strominger, \plt  B244
(1990) 393.}
\ldf\FMTV{
S. Ferrara, N. Magnoli, T. Taylor and G. Veneziano, \plt
 B245 (1990) 409.}
\ldf\Font{A.\ Font, {\it Periods and Duality Symmetries in \cy\
 Compactifications}, preprint UCVFC/DF-1-92.}
\ldf\FILQ{A.~Font, L.E.~Ib\'a\~nez, D.~L\"ust and F.~Quevedo,
\plt  B245 (1990) 401.}
\ldf\Forsyth{A. Forsyth, {\it Theory of Differential Equations}, Vol.
4,
Dover Publications, New-York (1959).}
\ldf\FS{P.\ Fr\`e and P.\ Soriani, {\it Symplectic embeddings,
K\"ahler
geometry and automorphic functions: The Case of $SK(n+1) =
   SU(1,1) / U(1) \times SO(2,n) / SO(2) \times SO(n)$}, preprint
SISSA
 90/91/EP.}
\ldf\GGRS{For a review see
S.~J.~Gates, M.~Grisaru, M.~Ro\v cek and W.~Siegel,
{\it Superspace}, \brk Benjamin/Cummings, 1983.}
\ldf\GS{A.\ Giveon and  D.-J.\ Smit, Mod. \plt A {\bf 6}
No. 24 (1991) 2211.}
\ldf\GSW{For a review see M.~Green, J.~Schwarz and E.~Witten,
{\it Superstring Theory}, Cambridge University Press, 1987.}
\ldf\GVW{B. Greene, C.\ Vafa and N.P.\ Warner, \nup324 (1989) 371. }
\ldf\ILR{L.~Ib\~anez, D.~L\"ust and G.~Ross, \plt272 (1991) 251;
       L.~Ib\~anez and D.~L\"ust, CERN preprint CERN-TH.6380/92
(1992).}
\ldf\IN{L. Ib\'a\~nez and H.P. Nilles, \plt  169B
(1986) 354.}
\ldf\kaplunovskya{V. Kaplunovsky, Phys. Rev. Lett  55 (1985)
1036.}
\ldf\kaplunovskyb{V. Kaplunovsky, \nup307 (1988) 145.}
\ldf\kaplunovskyc{V. Kaplunovsky,Texas preprint UTTG-15-91}
\ldf\KL{V. Kaplunovsky and J. Louis,Slac-Pub}
\ldf\KP{C.~Kounnas and M.~Porrati, \plt
   191B (1987) 91.}
\ldf\krasnikov{N.V.~Krasnikov, \plt  193B (1987) 37.}
\ldf\KST{
 A.~Klemm, M.~G.~Schmidt and S.~Theisen,\
{\it Correlation functions
for topological Landau-Ginzburg models with $c\leq3$}, preprint
KA-THEP-91-09.}
\ldf\Kos{B.\ Kostant, Am.\ J.\ Math.\ 81 (1959) 973.}
\ldf\lerche{W.\ Lerche,
\nup 238 (1984) 582; W.\ Lerche and W.\ Buchm\"uller,
Ann.\ Phys.\ 175 (1987) 159.}
\ldf\LSW{W.\ Lerche, D.\ Smit and N.\ Warner, \nup372 (1992) 87.}
\ldf\LVW{W.\ Lerche, C.\ Vafa and N.P.\ Warner, \nup324(1989) 427.}
\ldf\louis{J. Louis, SLAC-PUB 5527.}
\ldf\LT{D.~L\"ust and T.~Taylor, \plt  B253 (1991)
335.}
\ldf\maassarani{Z. Maassarani, \plt273B (1991) 457.}
\ldf\MN{G.~Moore and P.~Nelson, Phys. Rev. Lett.   53 (1984) 1519.}
\ldf\morrison{D.~Morrison,
{\it \pf\ equations and mirror maps for hypersurfaces},
 Duke preprint DUK-M-91-14.}
\ldf\nilles{H.~P.~Nilles, \plt  180B (1986) 240.}
\ldf\NO{H. P.~Nilles and M.~Olechowski, \plt B248 (1990)
268.}
\ldf\NSVZ{V.I.~Novikov, M.A.~Shifman, A.I.~Vainshtein
   and V.I.~Zakharov, \nup260 (1985) 157.}
\ldf\ross{G. Ross, \plt  211B (1988) 315.}
\ldf\seiberg{N.~Seiberg, \nup303 (1988) 206.}
\ldf\shenker{S.H.~Shenker, Lecture at
1990 Carg\`ese Workshop, Rutgers preprint RU-90-47;
A.~Dabholkar, \nup368 (1992) 293.}
\ldf\SV{M.A.~Shifman and A.I.~Vainshtein,
   \nup277 (1986) 456, \nup359 (1991) 571.}
\ldf\strominger{A.\ Strominger, \cmp133 (1990) 163.}
\ldf\taylora{T.R.~Taylor, \plt  164B (1985) 43.}
\ldf\taylorb{T.~Taylor, \plt  B252 (1990) 59.}
\ldf\TV{T. R.~Taylor and G.~Veneziano, \plt  212B
(1988) 147.}
\ldf\vafa{C.\ Vafa, Mod.\ Phys.\ Lett. A6 (1991) 337.}
\ldf\VW{E.\ Verlinde and N.P.\ Warner, \plt 269B (1991) 96.}
\ldf\weinberg{S.~Weinberg, Phys. Lett B91 (1980) 51.}
\ldf\witten{E.~Witten, \plt  149B (1984) 351
   , \plt  155B (1985) 151.}
\ldf\zwiebach{See for example B.~Zwiebach, MIT preprint MIT-CTP-1926
and references therein.}
%
%
%
%
%
\ldf\cubicF{E.\ Cremmer, C.\ Kounnas, A.\ van Proeyen, J.\
Derendinger,
S.\ Ferrara, B.\ de Wit and L.\ Girardello, \nup250 (1985) 385 \semi
   other cubic F-theories, de Wit xxxxxx?????}
\ldf\dfnt{B.\ de Wit and A.\ van Proeyen, \nup245 (1984) 89;
B.\ de Wit, P.\ Lauwers and A.\ van Proeyen, \nup255 (1985) 569;
E.\ Cremmer, C.\ Kounnas, A. \ van Proeyen, J.P.\ Derendinger,
S.\ Ferrara, B.\ de Wit and L.\ Girardello,  \nup250 (1985) 385.
}
\ldf\fetal{L.~Castellani, R.~D'Auria and S.~Ferrara,
    \plt B241 (1990) 57; Class.\ Quant.\ Grav.\ 1 (1990) 317;
R.~D'Auria,
S.~Ferrara and P.~Fr\'e, \nup359 (1991) 705.}
\ldf\cubicF{E.~Cremmer and A.~van Proeyen, Class. Quant. Grav.
{\bf 2} (1985) 445;
S.\ Cecotti, \cmp124(1989) 23;
B.~de Wit and A.~van Proeyen, {\it Special geometry, cubic
polynomials
and homogeneous quaternionic spaces},
CERN--preprint  TH.6302/91.}
%
%
%
\ldf\tntwo{E.\ Witten, \cmp118 (1988) 411, \nup340 (1990) 281; T.\
Eguchi and
 S.K.\ Yang, Mod.\ Phys.\ Lett.\ A5 (1990) 1693.}
\ldf\AMW{S.\ Cecotti and C.\ Vafa, \nup367 (1991) 359;
P.\ Aspinwall and D.\ Morrison, Duke
preprint DUK-M-91-12; Contributions of C.~Vafa and
E.\ Witten {\sl in} {\it Essays on Mirror Manifolds}, ed.
S.-T.~Yau, 1992, International Press, Hong Kong.}
\ldf\PFref{
A.\ Cadavid and S.\ Ferrara, \plt B267 (1991) 193;
W.\ Lerche, D.\ Smit and N.\ Warner, \nup372 (1992) 87;
D.~Morrison,
 Duke preprint DUK-M-91-14;
S.~Ferrara and J.~Louis, \plt 278B (1992) 240;
A.~Ceresole, R.~D'Auria, S.~Ferrara, W.~Lerche and J.~Louis,
CERN preprint CERN-TH.6441/92.}
%
%
%
\ldf\Ferrara{S.\ Ferrara and A.\ Strominger, {\it N=2 spacetime
supersymmtry and Calabi-Yau moduli space}, preprint CERN-TH-5291/89;
S.\ Cecotti, \cmp131(1990) 517;
A.\ Cadavid, M.\ Bodner and S.\ Ferrara, \plt B247 (1991) 25.}
%
%
%
%
\ldf\mirror{L.\ Dixon and D.\ Gepner, unpublished;
W.\ Lerche, C.\ Vafa and N.P.\ Warner, \nup324(1989) 427;
B.\ Greene and M.\ Plesser, \nup 338 (1990) 15;
P.\ Candelas, M.\ Lynker and R.\ Schimmrigk, \nup 341 (1990) 383.}
\ldf\mirrorreview{For a review see
B.\ Greene and M.\ Plesser, {\sl in} {\it Essays on Mirror
Manifolds}, ed.
S.-T.~Yau, 1992, International Press, Hong Kong,
and references therein.}
%
%
%
\ldf\gauginoc{For a review see for example H.~P.~Nilles,
Int. J. Mod. Phys. $\us{A5}$ (1990) 4199 and references
therein.}
\ldf\amati{D. Amati, K. Konishi, Y. Meurice, G. Rossi and
G. Veneziano, Phys. Rep.  162 (1988) 169.}
\ldf\dindrsw{J.P.~Derendinger, L.E.~Ib\'a\~nez and
  H.P.~Nilles, \plt  155B (1985) 65;
M.~Dine, R.~Rohm, N.~Seiberg and
  E.~Witten, \plt  156B (1985) 55.}
\ldf\effl{
A.~Font, L.E.~Ib\'a\~nez, D.~L\"ust and F.~Quevedo,
\plt  B245 (1990) 401;
S.~Ferrara, N.~Magnoli, T.~Taylor and G.~Veneziano, \plt
 B245 (1990) 409;
H.~P.~Nilles and M.~Olechowski, \plt B248 (1990)
268;
P.~Bin\'etruy and M. K. Gaillard, \nup358 (1991) 121;
M.~Cveti\v c, A.~Font, L.E.~Ib\'a\~nez, D.~L\"ust and
F.~Quevedo, \nup361 (1991) 194.}
\ldf\TVY{G.~Veneziano and S.~Yankielowicz, \plt
   113B (1982) 231;
  T.R.~Taylor, G. Veneziano and S.~Yankielowicz, \nup
   B218 (1983) 493;
T.R.~Taylor, \plt  164B (1985) 43.}
\ldf\hiddenmatter{D.~L\"ust and T.~Taylor, \plt B253 (1991)
335;
B.~de Carlos, J.~A.~Casas and C.~Mu\~noz, Phys. Lett B263 (1991)
248, CERN preprint CERN-TH.6436/92.}
\ldf\twogaugino{L.~Dixon, talk presented at the A.P.S. D.P.F.
  Meeting, Houston, 1990, SLAC-PUB 5229;
J.~A.~Casas, Z.~Lalak, C.~Mu\~noz and G.G.~Ross, \nup347 (1990) 243;
T.~Taylor, \plt B252 (1990) 59;
L.~Dixon, V.~Kaplunovsky, J.~Louis and M.~Peskin, unpublished.}
%
%
%
\ldf\wnrt{E.~Martinec, \plt 171B (1986) 189;
 M.~Dine and N.~Seiberg, \prl 57 (1986) 2625;
 J.~Atick, G.~Moore and A.~Sen, \nup307 (1988) 221;
 O.~Lechtenfeld and W.~Lerche, \plt 227B (1989) 375.}
\ldf\JJW{I.~Jack and D.~R.~T.~Jones, \plt258 (1991) 382;
P.~West, \plt258 (1991) 375.}
\ldf\frenormalization{M.A.~Shifman and A.I.~Vainshtein,
   \nup277 (1986) 456;
H.~P.~Nilles, \plt  180B (1986) 240;
I.~Antoniadis, K.~Narain and T.~Taylor, \plt B267 (1991) 37.}
%
%
%
%
\ldf\threegen{See for example
B.~Greene, A.~Lutken and G.~Ross, \nup325 (1989) 101;
G.~Ross, CERN preprint TH-5109/88 and references therein.}
%
%
%
\ldf\ellisetal{For a review see J.~L.~Lopez and D.~V.~Nanopoulos,
Texas preprint CTP-TAMU-76/91 and references therein.}
%
%
%
\ldf\danomaly{
J.~Louis, SLAC-PUB 5527;
J.~P.~Derendinger, S.~Ferrara, C.~Kounnas and F.~Zwirner,
CERN preprint CERN-TH.6004/91-REV;
G.~Cardoso Lopes and B.~Ovrut, \nup369 (1992) 351;
V.~Kaplunovsky and J.~Louis, SLAC-PUB to appear.}
\ldf\ILRE{I.~Antoniadis, J.~Ellis, S.~Kelley and D.~V.~Nanopoulos,
\plt B272 (1991) 31;
L.~Ib\'a\~nez, D.~L\"ust and G.~Ross, \plt B272 (1991) 251;
       L.~Ib\'a\~nez
       and D.~L\"ust, CERN preprint CERN-TH.6380/92 (1992).}
\ldf\DKLANT{L.~Dixon, V.~Kaplunovsky and J.~Louis, \nup355 (1991)
649;
I.~Antoniadis, K.~Narain and T.~Taylor, \plt B267 (1991) 37.}
%
%
%
\ldf\schell{See {\it Superstring Construction}, Current Physics
  Sources and Comments, Vol. IV, ed. A.~Schellekens
  (North-Holland, 1989).}
\ldf\twotworeview{For a review see
L. Dixon, {\sl in} Proc. of the 1987 ICTP Summer
Workshop in High Energy Physics, Trieste, Italy, ed.~G.~Furlan et
al.;
D.~Gepner, {\sl in} Proc. of the 1989 ICTP Spring School on
Superstrings,
Trieste, Italy, ed.~M.~Green et al.;
B.~Greene, Lectures at the ITCP
Summer School in High Energy Physics and Cosmology, Trieste, Italy,
1990,
and references therein.}
\ldf\cargese{See for example Proceedings of Carg\`ese Workshop on
Random Surfaces, Quantum Gravity and Strings, Carg\`ese, May 1990.}
%
%
%
\ldf\dualityreview{For a review see S.~Ferrara and S.~Theisen, {\sl
in}
Proc. of the Hellenic Summer School 1989, World Scientific;
D.~L\"ust, CERN preprint
CERN-TH.6143/91;
J.~Erler, D.~Jungnickel, H.~P.~Nilles amd M.~Spali\'nski, MPI
preprint
MPI-Ph/91-104
and references therein.}
%
%
%
\ldf\BH{See for example
Proceedings of the  Spring School on String Theory and Quantum
Gravity, ITCP, Trieste, April 1992.}
%
%
%
\ldf\CFGP{E.~Cremmer, S.~Ferrara, L.~Girardello and A.~van Proeyen,
\nup212 (1983) 413.}
%
%
%
\ldf\kln{S.~Kalara, J.~Lopez and D.~Nanopoulos, preprint
CTP-TAMU-46-91.}
\ldf\howi{Y. Hosotani, \plt  126B (1983) 309;\brk
 E. Witten, \plt  126B (1984) 351.}%
\ldf\thooft{G.~`t~Hooft, \nup72 (74) 461.}
\ldf\aeln{I.~Antoniadis, J.~Ellis, A.B.~Lahanas and
  D.V.~Nanopoulos, preprint CERN-TH.5604/89.}
\ldf\gs{
  J.~Scherk and J.H.~Schwarz, \nup81 (1974) 118;\brk
  D.~Gross and J.~Sloan, \nup291 (1987) 41.}
\ldf\bfq{C.~Burgess, A.~Font and F.~Quevedo, Nucl.
   Phys.  B272 (1986) 661.}
\ldf\ginsparg{P.~Ginsparg, \plt  197B (1987) 139.}
\ldf\ellis{
J.~Ellis, P.~Jetzer and L.~Mizrachi, \plt  196B (1987) 492
, \nup303 (1988) 1.}
\ldf\wb{J.~Wess and J.~Bagger, {\it Supersymmetry and
  Supergravity} (Princeton Unversity Press, 1983);\brk
and preprint JHU-TIPAC-9009, June 90.}
\ldf\PrandT{J.~Preskill and S.~Trivedi,
  \nup (Proc. Suppl.)  1A (1987) 83.}
\ldf\witind{E.~Witten, \nup202 (1982) 253.}
 \ldf\rv{G.C.~Rossi and G.~Veneziano, \plt  138B
    (1984) 195.}
\ldf\grisaru{M.T.~Grisaru, B.~Milewski and D.~Zanon,
  in {\it Supersymmetry and Its Applications -- Superstrings,
  Anomalies and Supergravity}, eds. G.W.~Gibbons, S.W.~Hawking and
  P.K.~Townsend (Cambridge Univ. Press, 1986).}
\ldf\cdg{R.~Jackiw and C.~Rebbi, Phys. Rev. Lett.  37 (76)
   132;\brk
  C.G.~Callan, R.F.~Dashen and D.J.~Gross,
   \plt  63B (76) 334.}
\ldf\serre{see for example J.-P. Serre, {\sl A Course in Arithmetic},
Springer-Verlag 1973, New York.}
\ldf\jones{D.R.T.~Jones, \nup87 (75) 127.}
\ldf\konishi{T.E.~Clark, O.~Piquet and K.~Sibold, \nup
   B159 (79) 1;\brk
   K.~Konishi, \plt  135B (1984) 439.}
\ldf\bganom{P.~Bin\'etruy and M.K.~Gaillard, \plt
   232B (1989) 83.}
\ldf\bq{C.P.~Burgess and F.~Quevedo, Phy. Rev. Lett. 64 (1990)
2611.}
\def\cernout{
\def\pubnum{\hbox{CERN-TH.6492/92}}
\def\pdate{
\hbox{CERN-TH.6492/92}
\hbox{May 1992}
}
\titlepage
\vskip 2cm
\title{Recent Developments in Superstring Phenomenology}
\author{Jan Louis}
\CERN
\vskip 1.5cm
\abstract{
Recent developments in superstring phenomenology are summarized
on a non-technical level. }
\vskip 3cm
\centerline
{\it Talk presented at the XXVIIth Rencontre de Moriond}
\centerline{\it on
Electroweak Interactions and Unified Theories.}
         }
%
\cernout
\endpage
Superstring theory \GSW\  is still today the only candidate theory
incorporating a consistent quantum  gravity.
That is, it provides a prescription of how to
sensibly compute loop corrections
to graviton--graviton scattering. Unfortunately, these corrections
are
far too small to be measured in the near future and hence this
aspect of the theory cannot be verified directly at present.
However, superstring theory also accommodates a spin 1 gauge
theory with
chiral fermions in the fundamental representation, which makes it a
candidate for a theory of all known interactions.\foot{In this
talk I exclusively focus on the so-called `heterotic string', which
is
phenomenologically the most promising string theory.}
Clearly, it should
be possible to check the validity of this assertion. To do so, one
needs to extract the predictions of string theory at the
experimentally accessible energy scale, the weak scale $\Mw$.
Being a consistent quantum gravity the characteristic scale of
string theory is the Planck mass
($\Mpl \sim 10^{19}$ GeV) and thus the physics at $\Mw$
is described by some low-energy effective field theory.
At first sight, this effective field theory resembles rather
closely (some extension of) the Standard Model (SM). It has a gauge
group which comfortably includes $SU(3)\times SU(2)\times U(1)$ and
`roughly' the right spectrum: family replication of light
chiral fermions in the fundamental representation as well as
Higgs-like states inducing gauge symmetry breaking.
Only when one tries to make this
low-energy limit more precise does one encounter a number of
technical
and phenomenological difficulties.
This is (partly) due to
the fact that string theory is only known
in a rather incomplete fashion: as a perturbative expansion.
Consequently we have at present no handle on its
non-perturbative properties and the physical effects hidden
in this regime.
Thus it might well be that once non-perturbative effects are taken
into account all phenomenological shortcomings disappear.
Being so close to
having a consistent theory of
all interactions with a light spectrum not unlike
the SM we hesitate to
abandon string theory
 too quickly, even if some of the more
detailed predictions
are not yet exactly matched in the particle world as
we observe it.
After all, the fermion masses agree
to leading order (they are zero), and  the subleading corrections
are a tiny effect with respect
to $\Mpl$
and require an enormous computational precision.
There are two other aspects that I would like to mention
in favour of string theory.
Firstly, string theory does have the potential to answer some
of the questions we have (so far) no way of addressing in the context
of the SM. In principle it explains the origin of the gauge group
and the fermion mass spectrum.\foot{Being a consistent quantum
gravity it also teaches us about
space--time and its singularity structure \BH.}
Even if we are not yet able to obtain the answers from
string theory, it seems a major advance to be able to formulate
these questions in a physical theory.
Secondly, string theory does inspire our imagination about possible
physics `beyond the SM'. Some of the  physical effects I mention
below could also occur in any other theory operating at some
high-energy scale.

In this talk I describe
some of the technical and phenomenological problems
of superstring theory, and
indicate the recent propositions of how non-perturbative physics
might evade them.
Let me first outline how one obtains the effective field theory
which is the low-energy limit of string theory.
The particle
spectrum
consists of a finite number of light fields (compared
with $\Mpl$) and an infinite number of
heavy states whose masses are
proportional to $\Mpl$.
The low-energy effective
Lagrangian depends only
on the light fields (these will be identified with the quarks and
leptons)
and the heavy modes contribute to their
effective interactions.
This process
of `integrating out' the heavy fields
depends on which ground state (or vacuum) of the string
theory has been chosen. For every ground state there can be
a different effective Lagrangian, a different low-energy
spectrum with different effective couplings.\foot{In the terminology
of `model building' each string ground state corresponds to a
different
`model'.}
Unfortunately, there exists an
enormous vacuum degeneracy of every
known string theory. That is, there are a large number of consistent
classical ground
states and there is  no physical mechanism  known at present
which distinguishes between them.\foot{ In fact, one is not even able
to enumerate or classify the space of all consistent string vacua.}
As I already indicated
it is commonly believed that a mechanism for lifting this degeneracy
is hidden in the
non-perturbative structure of string theory.
This state of affairs forces two
distinctly different strategies onto us. On the one hand, we need
to understand the non-perturbative properties of string theory.
Here we have
seen a lot of progress in simpler `toy' string theories where the
non-perturbative properties appear to be
 under control \cargese.
Hopefully further progress in this area will occur.
In this talk I will focus on
the other, more phenomenological
strategy and ask if there exists
any classical ground state which contains the SM at the weak scale.
Instead of randomly choosing a string ground state, one
selects instead those which have a chance of containing the SM.
Even in this phenomenological approach it would be a major
success if we were able to compute the top-quark mass along with all
the
other fermion masses. A vacuum that reproduces all measured
parameters of the SM clearly has some predictive power, since
it would tell us what kind of
`new physics' we can expect, at what scale.
Of course  this second approach
makes the assumption that string theory is
weakly coupled at $\Mpl$ so that we can trust
this perturbative analysis.
Indeed, we know that a theory  unifying  the
known interactions is
 weakly coupled at its characteristic scale
(which is $\Mpl$ in the case of string theory).
In practice one invokes a self-consistent analysis in which
weak coupling is assumed and then shown to arise.

In the process of selecting a `promising' vacuum one first chooses
a string ground state which corresponds to an
effective field theory in 4 space--time dimensions.\foot{This is
just a subset of all ground states and
many exist in dimensions other than
four. The hope is that string theory chooses dynamically a \fd\
vacuum over the others, which would explain why we live in four
rather than any other number of dimensions.}\
In addition one
usually requires it to be  supersymmetric at the Planck
scale. The reason for this is twofold,
non-supersymmetric vacua are often unstable at the loop level.
Secondly,
it is difficult to understand how a theory which operates at such
different energy scales can exist without supersymmetry.
(This is usually called the hierarchy problem.)
The requirement of four-dimensional space--time supersymmetry
constrains the couplings of the \el\ very strongly. In fact,
in the two-derivative approximation, they can be
expressed in terms of three arbitrary scalar functions:
the \K\ potential $K$, the superpotential $W$, and the gauge kinetic
function $f$ \CFGP. The $K$
governs the kinetic terms of the fields
($g_{ij}=\del_i\del_j K$,
$g_{ij}$ being the $\sigma$-model metric)
and  is thus important for
obtaining the proper field-normalization.
The effective
scalar potential $\veff$,
and thus the Yukawa couplings which eventually
determine the fermion masses, are expressed in terms of   $W$.
Finally, the
gauge couplings and the $\Theta$-angle
are encoded in $f$. In some sense,
string phenomenology is about computing $K,W,f$ for all vacua and
to all orders
in string perturbation theory, and if possible also
non-perturbatively.
Here one is helped by various non-renormalization theorems initially
proved in supersymmetry \GGRS\
and (partially) extended to string theory.
The superpotential
$W$ is not renormalized to all orders in string perturbation
theory \wnrt \foot{
A caveat to this theorem at two
\nrf{\JJW}
loops with massless fields present has been pointed out in
refs.~\refs{\GGRS,\JJW}.} whereas
the gauge kinetic function $f$ is believed to receive corrections
only at one loop \frenormalization.

Let me now briefly describe how $K,W$ and $f$ are
computed.
The motion of the string itself is described by a  two-dimensional
(conformal) field theory (CFT) and the effective couplings are
encoded in the
correlation functions of this CFT.
 The two requirements of a four-dimensional string vacuum,
 which is also
 $N=1$ space--time supersymmetric, translates into a condition
on the CFT. It turns out that
one needs a specific class of
CFTs which are  $N=2$ supersymmetric (in two dimensions)
and have a so-called central charge $c=9$.
The correlation functions of such a CFT have been studied
\nrf{\twotworeview \threegen \ellisetal}
extensively and were computed for a number of
examples \refs{\twotworeview{--}\ellisetal}.
In calculating $K,W,f,$ there have been two basic strategies. On the
one hand, particularly promising string vacua have been analysed in
detail \doubref\threegen\ellisetal.
On the other hand more generic properties,
common to all or a large class of vacua, have been uncovered
\twotworeview.
Unfortunately,
for most string vacua we are currently not able to calculate
the relevant correlation functions.
Therefore, it could well be that for the vacuum  which contains the
SM
we cannot compute any couplings.

Recently, there has been some progress in the technique of
computing the effective
Lagrangian.
For a particular string vacuum (a compactification on a
specific \cy\ threefold) some of its couplings have been
computed \CDGP\ (at the string tree level)
without  relying on  the underlying CFT, but instead employing
methods
of algebraic geometry and the recently
observed mirror symmetry \mirrorreview.
Even though the general lesson
behind this specific example is not yet completely understood,
it seems that
the couplings can be
computed without knowing
the CFT correlation functions.
Instead, the necessary information is encoded in the so-called
\LG\ potential from which one derives a set of coupled partial
differential equations (\pf\ equations) \PFref.
The effective couplings are then
obtained from the solutions of these
differential equations. Similarly, some of the couplings
also arise  as correlation functions of some appropriate
topological field theory which is a `twisted' (and much simpler)
 version of the original
CFT \AMW.
These new insights
allow us to extend the
computation of the \el\ to a new class of vacua and
in a rather simple way.

Another property of string theory has proved useful in constraining
the low-energy couplings.
 Most string vacua are believed to display a discrete
symmetry termed `duality' \dualityreview. It appears as a symmetry of
the string partition function and is
due to the presence of so-called
winding states, which arise as a consequence
of the one-dimensional nature of the string.
This symmetry has no field theory
analogue and is a truly `stringy' property of the theory.
The explicit form of the symmetry group is so far only known
for a few vacua but there (in conjunction with supersymmetry)
it proved very powerful in constraining the couplings
\nrf{\FLST\effl} \refs{\dualityreview{--}\effl}.
(The methods employed in \CDGP\ also allowed the
determination of the duality group.)
Apart from this technical merit this symmetry also implies an
interesting
physical phenomenon. Under the symmetry operation, small and large
distances are being identified, which indicates the existence of a
minimal length in nature.

So far I mentioned some of the technical problems (and the
recent improvements) in computing the low-energy effective Lagrangian
or equivalently $K,W,f$.
Let me now turn to some of the
phenomenological
problems which are common to all \fd\ supersymmetric
string ground states.
Generically, the gauge group $G$ in string theory is a product
of factor groups $G_a$
$$
G= \prod_a G_a    \, .
\eqn\ggroup
$$
An example is the well-known $E_8\times E_6$ but there are many vacua
with other gauge groups.\foot{However, the rank of $G$ cannot be
arbitrarily large, string theory puts an upper bound on it.}
 Still, string theory chooses the same (unified)
gauge coupling for every factor.
Thus, there is unification in the sense of a universal
gauge coupling. However, this does not imply a single gauge group
as in conventional GUT theories.\foot{
{}From the string point of view, there seems to be almost no reason
left for considering a conventional GUT theory.}
Furthermore, the unification
occurs at a different scale, the Planck scale.\foot{
In ref.~\ILRE\ it has been shown under what conditions this can be
consistent with the LEP precision measurements and, vice versa, what
conditions the LEP measurements put on the light spectrum of string
theory.}
In addition, this gauge coupling is
dynamical, that is, it
arises as the vacuum expectation value (VEV) of a scalar field
$\phi$ called the dilaton
$$
1/g_a^2 = \vev{\phi}  \, ,
\eqn\dilaton
$$
It can be shown that $\phi$ is a flat direction of the effective
potential $\veff$ to all orders in string perturbation theory.
Thus $\vev\phi$ is undetermined and a free parameter of the theory,
in contradiction with the SM.\foot{
There are other scalar fields in the spectrum which share the
property of being a flat direction of $\veff$.
These fields are called
moduli and their VEVs parametrize  continuously connected
families of string vacua. However,
they are not related to the tree-level
gauge couplings.}
This is the first `generic' problem.

If one chooses supersymmetric string vacua this supersymmetry
has to be broken somewhere between the \ps\ and the \ws.
The reason for choosing a supersymmetric vacuum in the first place
was the hierarchy problem. However, this requires not only
supersymmetry at $\Mpl$ but also that it be broken at some
intermediate scale around $10^{10}$--$10^{11}$ GeV.
It turns out to be surprisingly difficult to conceive a mechanism
which induces
such a hierarchical supersymmetry breaking
(second problem).

Even though in superstring theory
the cosmological constant is zero before supersymmetry breaking
one needs a  mechanism which keeps it at zero after
the breaking (third problem).

As I already indicated,
it is believed that the solution of these generic problems
lies in the non-perturbative regime of string theory.
This can only be analysed
under the assumption that the dominant
non-perturbative effects
are field theoretic in nature. Those are certainly contained in any
`stringy' \np\ effect. The real assumption is that they are
dominant.

Any non-perturbative effective potential arising in field theory
displays a dependence on the gauge coupling constant which is
of the form\foot{
The non-perturbative effect encountered in some toy string theories
are of order $e^{1/g}$ \shenker. It is important to further
investigate those effects.
}
$
\veff\sim e^{-1/g^2} \, .
$
A popular example for such a $g$-dependence is generated by
gaugino condensation.
This was already investigated in the context of supergravity in the
early 80's \gauginoc\ and first applied to string theory in
ref.~\dindrsw.
One assumes a so-called `hidden sector' (no renormalizable couplings
with the observable sector),\foot{They appear in many
string vacua, the $E_8$ in $E_8\times E_6$ is only one example.}
which consists
of an asymptotically free
non-Abelian gauge theory,  weakly coupled at $\Mpl$ but which becomes
strongly
coupled at some lower
\def\lc{\Lambda_c}
scale $\lc$. In such a theory the gauginos
(supersymmetric partners of the gauge bosons)
condense and
induce an effective potential of the form
(at leading order)
$$
\veff \sim \,  \sum_a  c_a
 {\rm exp}\left(- {24 \pi^2 \over b_a  g_a^2} \right)    \, ,
\eqn\lcestimate
$$
where the sum runs over the different factors of the hidden
gauge group,
$b_a$ are the 1-loop coefficients of the $\beta$ function, and
$c_a$ are some constants.
The field dependence
of $1/g_a^2$ [eq.~\dilaton]
enters $\veff$ and generates a potential for $\phi$.
Unfortunately, the minimum of this potential
occurs at
$\vev\phi=\infty$ ($g^2=0$) corresponding to a free-field theory.
This unacceptable state of affairs can change, once
loop corrections to gauge couplings are taken into account.
In
general, below $\Mpl$, the running gauge couplings are given by
$$
{1 \over g_a^2(p^2)} = \phi + b_a \ln {\Mpl\over p^2} + \Delta_a (T)
\, ,
\eqn\loopg
$$
where
$\Da$ are the (infra-red finite) threshold corrections \weinberg.
They arise from integrating
out the heavy fields and are generically
gauge-group-dependent. Thus they destroy the universality
of the tree-level gauge coupling.
If the masses of the heavy fields depend on the VEV of a light scalar
field $T$ (Higgs or modulus),
 these threshold corrections can induce further field dependence
 (in addition to
the dilaton)
into the gauge couplings and via
\lcestimate\ also into
the effective potential.

It was observed
in ref.~\krasnikov\ and further investigated in
ref.~\twogaugino\ that
a hidden sector of two gauge groups with very closely
matched $\beta$ functions (e.g. $SU(N)\times SU(N+1)$)
and a difference $\delta$ ($\delta = {\Delta_{N+1} \over N+1} -
{\Delta_N \over N}$)
in the field-independent (constant) threshold corrections might
stabilize the dilaton VEV.
Indeed, minimization of eq.~\lcestimate\ generates
a  VEV  whose size is controlled by large $N$ and to leading
order in $N$ is given by
$
\re S \, = \, {N^2 \delta \over 16 \pi^2}
$
(for example $SU(9) \times SU(10)$ with
$\delta  = 4$ leads to $\re S = 2.1$ or $\alpha_{\rm GUT}
\sim 1/25$).
Thus, by taking into account the constant
threshold corrections, it is
possible to fix the dilaton VEV at a phenomenologically acceptable
value.
Unfortunately, without any $T$ dependence of $\Da$,
supersymmetry is unbroken at this minimum.

The computation of $\Da$  in string theory
can be performed
explicitly when the complete mass spectrum is known \kaplunovskyb.
In ref.~\DKLANT\
the field dependence of $\Da(T)$ was computed
for a particular class of string vacua (orbifolds) and the
dependence on a specific set of massless modes $T$ (untwisted moduli).
It was found that
$$
\Da(T) = A_a \left(\ln (|\eta(T)|^4) + \ln (T+\bar T)\right) \, ,
\eqn\forbi
$$
where $A_a$ are some constants and $\eta$ is the Dedekind
$\eta$ function.
The functional form of $\Da(T)$ strongly depends on the
vacuum chosen. However,
the computation of $\Da(T)$ can be somewhat simplified by observing
that  the two terms in eq.~\forbi\
correspond to the
contribution of the heavy  and light states running in the
loop \danomaly.
The contribution of the light states
can be calculated rather easily from the tree-level
couplings alone. By employing
symmetry arguments (for example the duality mentioned above)
this might be
sufficient to infer the full structure of $\Da$.\foot{
For the example
of the \cy\ manifold referred to earlier, this procedure
has been carried out in ref.~\dixonberkeley.}
Clearly it is important to understand the field dependence
of $\Da$ for a larger class of vacua and also the dependence on other
(non-moduli) fields.

It was shown in ref.~\effl\ that once the dilaton is fixed
supersymmetry can be broken
by including field-dependent $\Da(T)$.
Now the potential has to be minimized in a multidimensional
field space, which can have non-supersymmetric minima.
This computation was performed for the particular class of vacua
where $\Da$ is given by eq.~\forbi.
Even though this is a small subset of all vacua,
it is encouraging that supersymmetry breaking occurs,
suggesting that it might happen under much more general
conditions.
As I already indicated, supersymmetry breaking alone is not enough:
it has to occur at the right scale. By considering
hidden gauge groups
containing some `hidden matter', the generated hierarchy can be
improved \hiddenmatter.
The second important aspect is that
the same mechanism also determines the
VEV of the modulus $T$. This
leads to a (partial) lifting of the vacuum degeneracy.

However, we are left with a number of shortcomings.
First of all, in almost all models a cosmological constant
is induced after supersymmetry breaking.
This prohibits a further analysis of soft supersymmetry-breaking
terms and more detailed low-energy phenomenology.
Secondly, the two mechanisms proposed above, the fixing of $\vev\phi$
and supersymmetry breaking,
have not been put
together in a realistic string vacuum. Thus we are still
at the stage of suggesting possible scenarios of how the string
might choose to break supersymmetry and fix the gauge coupling
constant.
The  recent developments are very encouraging, though.

Finally, let me speculate about another interesting physical
effect related to $\Da$. It is likely that $\Da$ also
depends on scalar
fields charged under the observable gauge group. When minimizing
$\veff$, this might be another mechanism to
induce gauge symmetry breaking.

To summarize,
I indicated some recent technical progress in computing the \leel.
This is an important step in comparing the Lagrangian with the SM.
Furthermore, by including threshold corrections of the gauge
couplings
into the non-perturbative effective potential, one is able to
fix the dilaton VEV at a phenomenologically acceptable value.
The field dependence of these corrections can also induce
supersymmetry
breaking. Unfortunately, we have made little progress on the problem
of the cosmological constant.

\ack
I would like to thank the organizers of this conference for
creating an unusually stimulating atmosphere, and D.~L\"ust for
useful comments on the manuscript.

\refout
\end